\documentclass{PoS}
\usepackage{lineno}
\title{Cosmic Ray Extremely Distributed Observatory: Status and perspectives of a global cosmic ray detection framework}

\ShortTitle{Cosmic Ray Extremely Distributed Observatory: Status and perspectives}

\author{\speaker{Dariusz G\'ora}$^{,a}$, 
Kevin Almeida Cheminant$^{a}$,
David E. Alvarez Castillo$^{e}$,
Dmitriy Beznosko$^{b}$,
Niraj Dhital$^{a}$,
Alan R. Duffy$^{c}$, 
Piotr Homola$^{a}$,
Konrad Kopa\'nski$^{a}$,
Peter Kovacs$^{d}$,  
Marta Marek, 
Alona Mozgova$^{h}$,
Vahab Nazari$^{a,e}$,
Michal Nied\'zwiecki$^{f}$, 
Wojciech Noga$^{f}$,
Katarzyna Smelcerz$^{a,f}$,
Karel Smolek$^{g}$,
Jaroslaw Stasielak$^{a}$,
Oleksandr Sushchov$^{a}$,
Dominik Ostrog\'orski$^{i}$,  
Krzysztof Rzecki$^{j}$, 
Krzysztof W. Wo\'zniak$^{a}$,
Jilberto Zamora-Saa$^{k}$
\\\
  
        E-mail: \email{Dariusz.Gora@ifj.edu.pl}\\
        $^{a}$ Institute of Nuclear Physics Polish Academy of Sciences, Radzikowskiego 152, Cracow, Poland\\
        $^{b}$ Harvard University, 1 Oxford Street, Cambridge, MA, USA \\
        $^{c}$ Centre for Astrophysics and Supercomputing, Swinburne University of Technology, Hawthorn, VIC 3122, Australia\\       
        $^{d}$ Institute for Particle and Nuclear Physics,Wigner Research Centre for Physics, Hungarian Academy of Sciences, H-1525 Budapest, Hungary \\
        $^{e}$ Joint Institute for Nuclear Research, Dubna, Russia  \\
        $^{f}$ Institute of Telecomputing, Faculty of Physics, Mathematics and Computer Science, Cracow University of Technology, Warszawska 24st 31-155 Cracow, Poland\\
        $^{g}$ Institute of Experimental and Applied Physics, Czech Technical University in Prague\\
        $^{h}$ Astronomical observatory of Taras Shevchenko National University of Kyiv\\
        $^{i}$ AGH University of Science and Technology, 30-059 Cracow, Poland\\
        $^{j}$ Cracow University of Technology, Warszawska 24st 31-155 Cracow, Poland \\
        $^{k}$ Universidad Andres Bello, Departamento de Ciencias Fisicas, Facultad de Ciencias Exactas, Avenida Republica 498, Santiago, Chile
        }

\abstract{The Cosmic-Ray Extremely Distributed Observatory (CREDO) is a project dedicated to global studies of extremely extended cosmic-ray phenomena, the cosmic-ray ensembles (CRE), beyond the capabilities of existing detectors and observatories. Up to date cosmic-ray research has been focused on detecting single air showers, while the search for ensembles of cosmic-rays, which may spread over a significant fraction of the Earth, is a scientific terra incognita. The key idea of CREDO is to combine existing cosmic-ray detectors (large professional arrays, educational instruments, individual detectors, such as smartphones, etc.) into a worldwide network, thus enabling a global analysis. The second goal of CREDO involves a large number of participants (citizen science!), assuring the geographical spread of the detectors and managing manpower necessary to deal with vast amount of data to search for evidence for cosmic-ray ensembles. In this paper the status and perspectives of the project are presented.}

\FullConference{36th International Cosmic Ray Conference -ICRC2019-\\
		July 24th - August 1st, 2019\\
		Madison, WI, U.S.A.}

\begin{document}

\section{Introduction}

The Cosmic-Ray Extremely Distributed Observatory (CREDO) aims at searching for the yet
not checked multi cosmic-ray signatures that are composed of many air showers and individual
particles arriving simultaneously to the Earth, so-called cosmic-ray ensembles (CRE).  
A good candidate for a CRE is a shower induced by an ultra high energy photon interacting close by the Sun or at least at some distance from the Earth.  
Such cascades, which we call preshowers~\cite{ps}, are produced as a consequence of  interaction of ultra high energy photons  (with energies larger than $10^{17}$ eV)  with solar or terrestrial  magnetic fields. Consequently, in case of an interaction close to the  Sun, by the time the cascade arrives at the Earth, it comprises several thousand photons and a few leptons with a peculiar spatial distribution to which CREDO will be  sensitive~\cite{niraj}. Another scenario for the production of CRE  lies in so-called {\it top-down} models. Among them, the
decay of long-lived super-massive particles (M$_{X}$ > $10^{20}$ eV)  may lead to a significant fraction of
UHE photons~\cite{dm} in cosmic ray flux and thus CRE.


The signatures of CRE might be spread over very large surfaces ($\sim$ 1000 km's) which might make them hardly
detectable by the existing detector systems operating individually. On the other hand, if these
detector systems operate under a planetary network, as proposed by CREDO, the chances for
detection of CRE will naturally increase. The components of CRE might have energies that
practically span the whole cosmic-ray energy spectrum. Thus, all the cosmic ray detectors working
in this range, beginning from smartphones (e.g. DECO, CRAYFIS, CREDO Detector) and pocket
scintillators (e.g. Cosmic Watch or CosmicPi), through numerous larger educational detectors and
arrays (e.g. HiSPARC, Showers of Knowledge, CZELTA) to the professional infrastructure that
receive or will receive cosmic rays as a signal or as a background (Pierre Auger Observatory,
Telescope Array, JEM-EUSO, HAWC, MAGIC, H.E.S.S., VERITAS, IceCube, Baikal GVD,
ANTARES Telescope, European Southern Observatory, other astronomical observatories,
underground observatories, accelerator experiments in the off-beam mode) could contribute to a
common effort towards a hunt for CRE.

The CREDO Collaboration, which is now an international collaboration of 23
institutions from 11 countries~\cite{credo}, apart from the involvement of physicists, will require a massive
participation of non-scientists. CREDO mission and strategy requires active engagement of a large number of participants, also non-experts, who will contribute to the project by using common electronic devices (e.g. smartphones). Such activities  will create a highly diversified, multiplanar and
multidimensional potential, including the potential to enlarge intellectual capacity of the whole
society.

\begin{figure}[t]
\center
\includegraphics[width=0.8\textwidth]{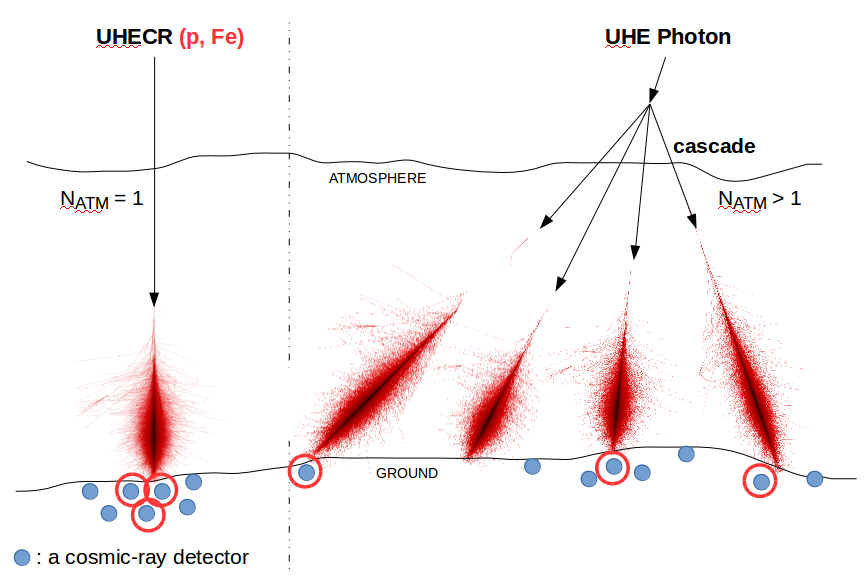}
\caption{Left: Typical strategy: search for one extensive air shower (EAS)  -- cascade of secondary particles initiated by a single high-energy cosmic ray; Right:
 Cosmic-Ray Ensembles: a novelty in cosmic-ray research and target for CREDO.}
\label{fig:2}
\end{figure}
\section{Concept,  methodology and CREDO status}


The scientific novelty of the CRE-oriented research is schematically illustrated in Figure~\ref{fig:2}.
The  novelty  is based on a general approach to large scale correlations of cosmic rays, with the
ambition to consider the widest possible range of scenarios that can be verified on Earth through an
observation of an ensemble of at least two particles or photons with a globally spread and
coordinated network of detectors. In addition to testing spatial correlations of particles arriving
simultaneously at the Earth, the general  CREDO strategy includes also searches for groups of spatially
correlated cosmic-ray photons that might arrive at the Earth at significantly different times, with
temporal dispersion of the order of minutes or more. Such phenomena have been reported in the
literature~(\cite{ev1,ev2}), but they have not been observed repeatedly until now. One of the
CREDO capabilities is either to confirm or exclude a connection character of these outlying
observations.

\begin{figure}[ht]
\center
\includegraphics[width=0.35\textwidth,height=7cm]{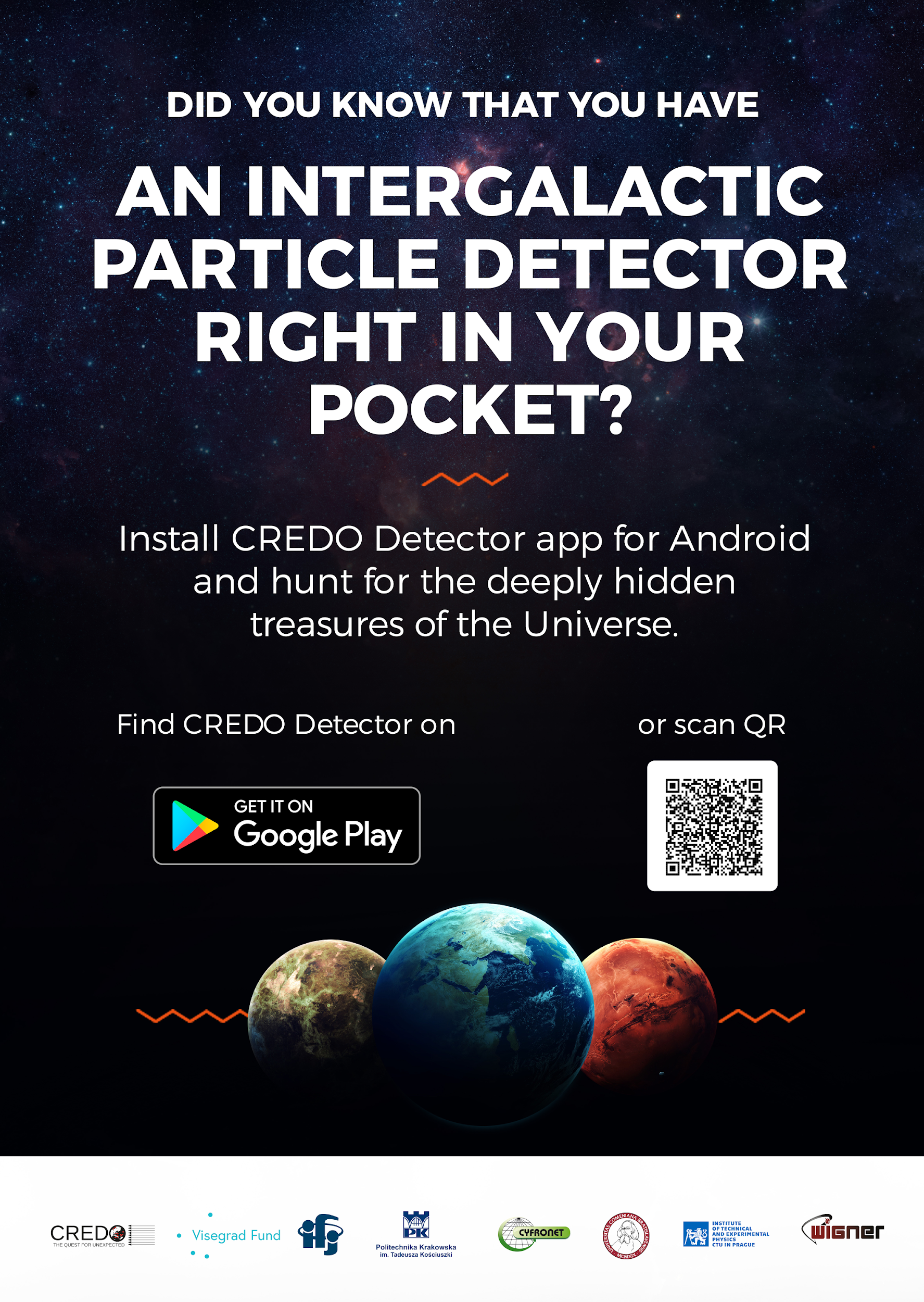}
\includegraphics[width=0.45\textwidth,height=7.1cm]{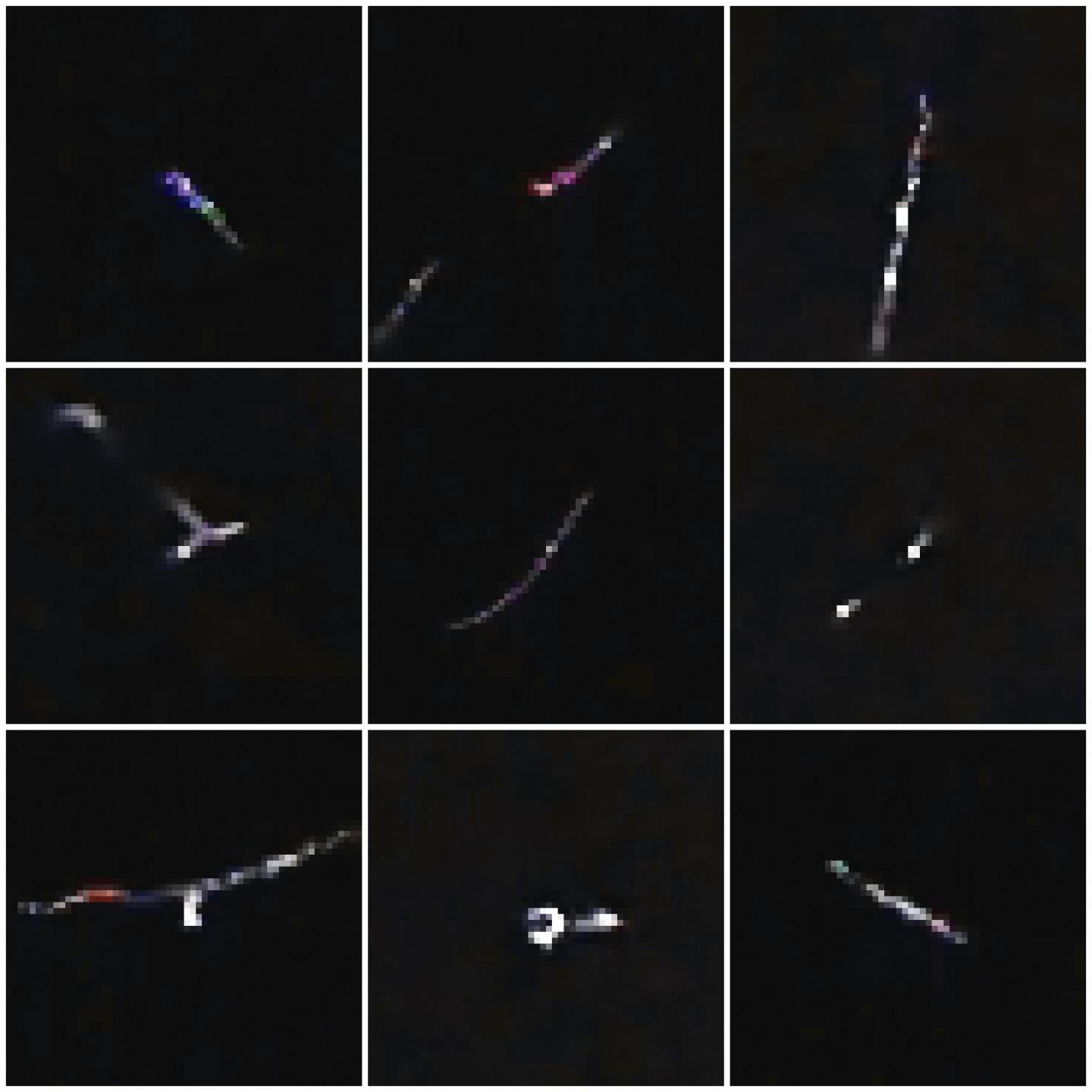}
\caption{ Left panel: The CREDO Detector app on Google Play; Right panel: Tracks of particles detected by the CREDO Detector app.}
\label{fig:4}
\end{figure}

Cosmic-ray
data are available everywhere on the planet, also inside buildings, they are  free, and they can be
acquired with the minimum detection effort in an easy reach of the public - through a mobile device
equipped with an application that turns it into a particle detector. To optimize the engagement scale,
we will use the already existing app, the CREDO Detector~\cite{detector} (see also Figure \ref{fig:4}), with the source code open under the
MIT license that gives a freedom of development driven even by a wide community, and unlimited
flexibility to implement all the features essential for the project.
The other unique feature of the app
is that it connects the user to the open server-side data processing, analysis, and visualisation
system~\cite{apicredo}, with open access API ready to receive data, also from detectors other than smartphones, and
to offer an open access to the data in close to real time.
 It is important to stress that all the CREDO related codes developed so far are available  on github repository~\cite{github} which enables an easy teaming up and continuation of the existing
projects, together with an active participation and engagement of non-experts, e.g. through
submitting issues. 

In near future  we are going to develop further the existing prototypes, supplement the data acquisition system with the iOS version of the
app, complementary detectors and data mining toolbox. We will also enrich the technical and
scientific content of the system with multi-level educational resources.

Since a smartphone with the CREDO Detector is capable of detecting particle track candidates (see
Figure  \ref{fig:4}  for examples) including both the cosmic radiation (e.g. the penetrating particles like muons) and
the local radiation (e.g. X-rays) it opens a door to the public engagement model which is based not only
on the data classification, like for most of the citizen science projects, e.g. in zooniverse.org, but also on
the data acquisition.  

\begin{figure}[ht]
\center
\includegraphics[width=0.8\textwidth]{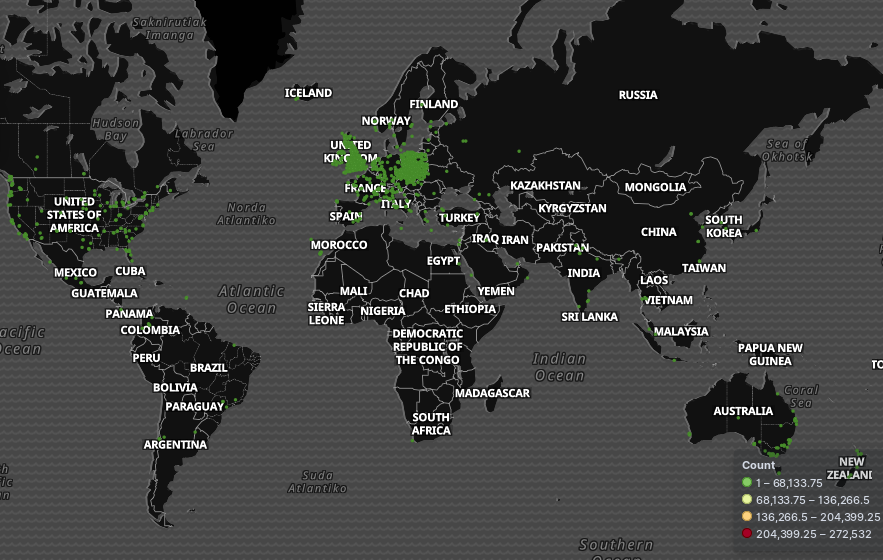}
\caption{ The map of CREDO user locations as of middle July 2019. More than 7500 users spread over the globe.}
\label{fig:5}
\end{figure}

In Figure \ref{fig:5} we show  the map of the CREDO user's locations, from which we can see that CREDO smartphones network already spreads over the planet.
The number of registered users, with at least one detection, in the middle of July  2019 was  7500  and  about 2 918 000 images were stored in the database. The  observing time  for all users  equates to 958 years searching for particles, which shows the large potential of such observations.

\begin{figure}[ht]
\center
\includegraphics[width=\textwidth]{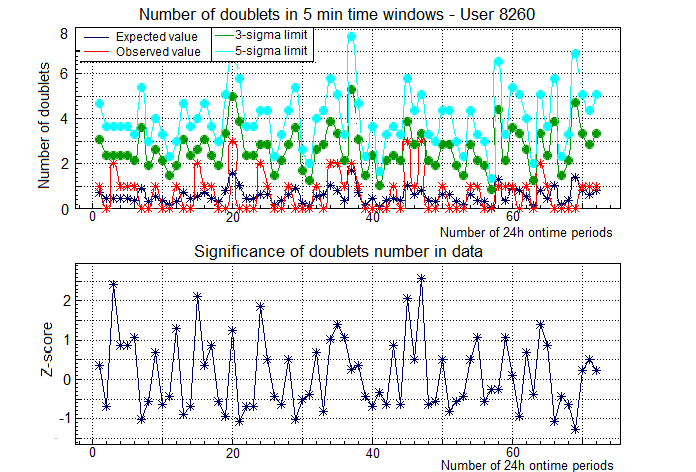}
\caption{Upper panel:  Number of doublets (two cosmic ray detections in an interval of 5 minutes) found in data (red) in every 24 hour period versus the expected value obtained from background simulations(dark blue). Number of doublets needed to obtain a 3 or 5 sigma effect are also given (green and light blue, respectively); Lower panel: Significance of number of doublets (two cosmic ray detections in an interval of 5 minutes) found in data in every 24 hour period.}
\label{fig:6}
\end{figure}

One of the developed tools of the CREDO infrastructure is the Quantum Gravity Previewer (QGP)~\cite{qgp} which
searches for clusters of time-correlated events in smartphones data. As an example, in Figure \ref{fig:6} we show one of the first results of the CREDO using this automated analysis of smartphones data. The plot show the distribution of doublets measured by an individual user in comparison to the expected number of such doublets. More precisely, the number of the expected doublets  for a fixed period is obtained  by using  the  so-called scrambled maps.
The scrambled map is a table containing randomly generated timestamps using a flat distribution, and with the number of events
equal to the number of timestamps in the user's data.  A significance parameter of a given doublet  is   calculated as described
 in~\cite{universe}. As we can see from the plot, for this individual user the number of doublets is compatible with the background expectations.

The unique scientific advantages that the smartphone cloud has in comparison to other detector
systems is the geographical spread and the public availability at no additional  investment. These two
features make the smartphone cloud a critically important and efficient component of the whole
CREDO system: if CRE typically spans surfaces much larger than those engaged even by the largest
 existing observatories, the only chance for detection might be to explore the potential of a global
network of detectors, such as CREDO, and the first step in this direction is what we have at hand:
smartphones.  The other important step is to complement the smartphone cloud with  more
advanced, but still affordable, detectors using different detection techniques. Such a supplementary effort will
be necessary to meet the standards of scientific excellence: even the most peculiar anomaly
observed with one technique only, e.g. by smartphone sensors, cannot be considered a real physical
phenomenon unless it is observed independently, with detectors of other types. Within this project
we will develop a range of  detectors based on the scintillation technique, that will have
open design and which will be advertised  primarily  among the most active teachers and the advanced
amateurs. The affordable scintillator detectors coupled to the smartphones via the central
computing system will form the core infrastructure for this project, to make it independent of
the time scale when the other potentially interested partners and big observatories join the global
cosmic ray effort.

 The CREDO Collaboration has also developed prototypes of the citizen science
workflows to be developed and to work within this project: Dark Universe Welcome~\cite{darkuniverse1}
 and Private Particle Detective~\cite{darkuniverse2}.
While the two workflows
are located on the Zooniverse platform, where participation of external developers is limited, the
ideas behind the workflows are transferable to other existing citizen science platforms with open
source codes -- thus our current prototypes can be considered showcases encouraging to build clones implementable within other open source citizen science platforms - to increase their flexibility and to benefit from contributions of external developers.

CREDO data needs to be stored  and secured from any unauthorized access. This task will be fulfilled by Cyfronet~\cite{cyfronet}, which has over 45 years of experience in providing High Performance
Computing and Mass Storage services for the Polish scientific community and has been engaged in
many projects which include experimental data preservation, such as LHC or ESA Sentinel. Data
gathered by the detectors will be stored and processed in the Cyfronet Data Center in Krakow, which
provides high availability of all necessary facilities, such as power and cooling, and ensures high level of
of physical security of the building and server rooms. Data will be backed up locally and also to
Cyfronet's remote Disaster Recovery location few kilometers away from the main DC and stored on
magnetic tapes, which provide high endurance and reliability for long-term storage purposes. This
assures, that the data is preserved and secured from any application or human error interference.

\section{Citizen science}
Citizen Engagement and Education in Science requires some initial curiosity from which
a strong motivation to go deeper into specific subjects can be built. This can not be achieved if the
public is treated as passive consumers of information from science only. Providing exhibitions,
events, lectures and articles on scientific discoveries is important, but does not ensure any further
engagement in science. This is possible if some follow-up activities are proposed, properly designed
to the audience. CREDO has a wide programme of such activities targeting persons of all ages and
backgrounds. In the case of the youngest kids their participation in events organized by CREDO
builds a positive image of science~\cite{citizen}. All persons using smartphones can be involved in measurements
of cosmic rays, no prior knowledge or special skills are required. The school students are
encouraged to become more active by taking part in a competition. Science enthusiasts can take part
in the analysis of collected data and in the development of tools needed for this. This variety of
engagement levels is  a compelling feature of the CREDO project.

As an example of such activities, 
we have launched a pilot team competition programme for schools and other educational organizations: the Particle Hunters League~\cite{liga},
and organized  a very successful first
CREDO workshop for Polish teachers, held in October 2018 in Krakow~\cite{credoweek}. 
There were 55 teacher participants and to date we
have nearly 60 school teams, and more than 1,200 individual pupils active in the Particle Hunter
League, which might mean that the teachers advertise the CREDO format among themselves -
which is the expected and desired effect. Based on this experience we are confident that the
CREDO educational format is sufficiently universal and adoptable in schools on large, global scales.
Thus we expect to achieve a very large educational impact once the CREDO programme is
 promoted properly.


\section{Other activities}

CREDO offers inter- and trans-disciplinary opportunities, and some of them might turn out to
be of vital everyday life importance for a large fraction of humanity. We list here a potential for
studying correlations of cosmic-ray flux with earthquakes based from a multi-messenger signal
received by a global network of detectors, including cosmic-ray sensitive devices operated now and
in the future by CREDO. An effort in this direction has a motivation in the literature~\cite{earth1,earth2,earth3,earth4}
 and is already being discussed on international fora~\cite{paris}.
The attempt to predict earthquakes with the help of the cosmic-ray information from a global
network exceeds state-of-the-art and paradigmatic thinking both in the cosmic-ray and geophysical
communities. 

Another interdisciplinary opportunity that cannot be missed in a complete description of the
CREDO ambitions is a global monitoring of risk from extensive air showers and, potentially, from
CREs. Average doses from cosmic ray radiation are known to be
significantly lower than those due to local and natural radiation, but it is also known that extensive
air showers  might potentially be harmful to biosystems, including humans. An adult human body located near the core
of an EAS induced by a primary cosmic ray of energy $10^{18}$ eV, or higher, is exposed to an energy
deposit equivalent to a cumulative dose of 40 milligrays (mGy), or larger, which has been recently
shown to be dangerous to humans~\cite{med1}. Given the cosmic ray flux at the energy of $10^{18}$ eV, 40 mGy or
larger, is applied in one shot to a significant fraction of the human population at least once during
an individual lifetime~\cite{med2}.

\section{Summary}
Pursuing the research strategy proposed in CREDO will have a large impact on astroparticle physics and possibly also on fundamental physics. If CRE are found, they could point back to the interactions at energies close to the Grand Unified Theories (GUT) scale.  This would give an unprecedented chance to test  experimentally for example  dark matter models.  If CRE are not observed it would valuably constrain the current and future theories. Apart from addressing fundamental physics questions  CREDO has a number of additional applications: integrating the scientific community (variety of science goals, detection techniques, wide cosmic-ray energy ranges, etc.), helping non-scientists to explore Nature on a fundamental but still understandable level. The high social and educational potential of the project gives confidence in its contributing to a progress in physics.

{\bf Acknowledgements:} This research has been supported in part by PLGrid Infrastructure. We warmly thank the staff at the ACC Cyfronet AGH-UST, for their always helpful supercomputing support. CREDO application is developed in Cracow University of Technology.

\end{document}